\documentstyle[12pt]{article}
\begin{document}
\newcommand{\vv}[1]{\stackrel{\rightharpoonup}{#1}}

\title{The combination of Wilson Loops into $J^{PC}$ Operators in Lattice
       Pure Gauge Theory}
\author{{Da Qing Liu$^2$, Ji Min Wu$^{1,~2}$ and Ying Chen$^2$}\\
        {\small $^1$CCAST(World Lab.), P. O. Pox 8730, Beijing 100080, China}\\
        {\small $^2$Institute of High Energy Physics, Chinese Academy of
                Sciences,}\\
        {\small P. O. Box 918-4, Beijing, 100039, P. R. China}}
\maketitle
\begin{center}
\begin{minipage}{5.5in}
\vskip 0.8in
{\bf Absract}\\
To calculate the mass of glueballs with quantum number $J^{PC}$ in
lattice
pure gauge theory by using Wilson loop operators, we discuss the
combination of Wilson loops
into irreducible representation of $O^{PC}$ group for
arbitrary-link Wilson loops. We present a general  computational
procedure for this combination which is suitable for any finite
group.
\end{minipage}
\end{center}
\vskip 1in
\noindent

\newpage

\section {Introduction}
To calculate the glueball mass spectra, one has to construct different
$J^{PC}$ glueball operators which act on vacuum to generate states with
certain $J^{PC}$ quantum number. These operators should be the correct
linear combinations of Wilson loops and transform as certain irreducible
representations $R^{PC}$ of group $O^{PC}$ if we also concern about space
parity P and charge conjugate C with cubic group $O$
on lattice. Berg and Billoire$^{\cite{s1}}$ has discussed
this problem and merged the problem of constructing $J^{PC}$ glueball
operators into the combinaton  of Wilson loops
with different lengths and types into irreducible representation $R^{PC}$ of
group $O^{PC}$. They also reveal the result of 4-link and 6-link
operators$^{[1]}$.

Therefore, the decomposition of the cubic group is the preliminary step to
calculate glueball masses$^{\cite{s7}}$. On the contrary,
since some representation matrices of the group $O^{PC}$ is as
large as $96 \times 96$, the calculation is lengthy and some time impossible by
hand. It is necessary to find a general and practically easy procedure
for the combination of arbitrary-link operators into $R^{PC}$. We would like
to introduce such a procedure here, which can be realized by computer
and is also suitable for any finite group.

Section 2 gives some appointments in our calculation and section 3 presents
our computational procedure. We present our results in section 4 and a short
summary is in section 5.
Appendix A shows some property of the cubic group. Appendix B is
some lengthy results from our calculation.

\section {Some Appointments}
We use L-component vector $(f_1,f_2,\cdots,f_L)$ $(f_i=\pm 1, \pm 2,\pm 3)$, to
represent L-link closed Wilson loop in which each component stands
for the space direction of the link along the Wilson loop.  For
instance, we denote a 4-link Wilson operator
$$
Tr(U_1(n)U_2(n+a\widehat{e}_1)U^{-1}_1 (n+a\widehat{e}_2)U^{-1}_2 (n))
$$
by $(1,2,-1,-2)$. Each vector forms a base of one representation. Therefore,
it is obvious that under C transformation:
$$(f_1,f_2,\cdots,f_L) \stackrel{C}{\rightarrow}
(-f_L,\cdots,-f_2,-f_1),$$
and under $M_g \in O^P$: $M_g(f_1,f_2,\cdots,f_L)=(M_g f_1,M_g f_2,\cdots,M_g
f_L).$

The cubic group $O$ has 24 elements which may be divided into five
classes
of conjugate elements: $E, C_2, C_3, C_4, C_2^2 $. We choose the
proper ones from each class of the conjugate elements which make
the directions  transformation as follows under their actions:
\begin{flushright}
$
\begin{array}{lllr}
1 \stackrel{C_2}{\rightarrow} -1, & ~~~2 \stackrel{C_2}{\rightarrow}-3, &
~~~3 \stackrel{C_2}{\rightarrow} -2; & ~~~~~~~~~~~~~~~~~~~~~~~~(1a)
\end{array}
$\\
$
\begin{array}{lllr}
1 \stackrel{C_3}{\rightarrow} -3,&~~~ 2 \stackrel{C_3}{\rightarrow}-1, &
~~~3 \stackrel{C_3}{\rightarrow} 2; & ~~~~~~~~~~~~~~~~~~~~~~~~~~~(1b)
\end{array}
$\\
$
\begin{array}{lllr}
1 \stackrel{C_4}{\rightarrow} -3,&~~~ 2 \stackrel{C_4}{\rightarrow} 2, &
~~~~~3 \stackrel{C_4}{\rightarrow} 1; & ~~~~~~~~~~~~~~~~~~~~~~~~~~~(1c)
\end{array}
$\\
$
\begin{array}{lllr}
1 \stackrel{C^2_4}{\rightarrow} -1,& ~~~~2 \stackrel{C^2_4}{\rightarrow}2,
&
~~~~~3 \stackrel{C^2_4}{\rightarrow} -3. &~~~~~~~~~~~~~~~~~~~~~~~~ (1d)
\end{array}
$
\end{flushright}
For generators$^{[2]}$~$C_3$ and $C_4$ of the cubic group $O$, we appoint
their representation matrices as:
\begin{flushright}
$
\begin{array}{llr}
 D^{A_1}(C_4)=1,~~~&D^{A_1}(C_3)=1;~~~~~~~~~~~~~~~~~~~~~~~~~~~&(2a)
\end{array}
$\\
$
\begin{array}{llr}
 D^{A_2}(C_4)=-1,~~~&D^{A_2}(C_3)=1;~~~~~~~~~~~~~~~~~~~~~~~~~~&(2b)
\end{array}
$\\
$
\begin{array}{llr}
 D^{E}(C_4)={1\over 2}\left(
                        \begin{array}{cc}
                              1 & 3\\
                              1 & -1
                        \end{array}
                      \right)
,~~~&
 D^{E}(C_3)=\frac{1}{2}\left(
                        \begin{array}{cc}
                         -1 & -3\\
                          1 & -1
                        \end{array}
                      \right)
          ;~~~~~~~~~~&(2c)
\end{array}
$\\
$
\begin{array}{llr}
  D^{T_1}(C_4)=\left(
                      \begin{array}{ccc}
                        0 & 0 & 1\\
                        0 & 1 & 0\\
                        -1 & 0 & 0
                      \end{array}
               \right)
          ,~~~&
  D^{T_1}(C_3)=\left(
                      \begin{array}{ccc}
                         0  & -1  &  0\\
                         0  &  0  &  1\\
                         -1 &  0  &  0
                      \end{array}
               \right)
          ;~~~~~~~~~~&(2d)
 \end{array}
$\\
$
\begin{array}{llr}
  D^{T_2}(C_4)=\left(
                      \begin{array}{ccc}
                        0  &  0  & -1\\
                        0  & -1  &  0\\
                        1  &  0  &  0
                      \end{array}
               \right)
          ,~~~&
  D^{T_2}(C_3)=\left(
                      \begin{array}{ccc}
                         0  &  -1 &  0\\
                         0  &  0  &  1\\
                         -1 &  0  &  0
                      \end{array}
               \right)
          ;~~~~~~~~~~&(2e)
 \end{array}
$
\end{flushright}
Including P and C, group $O^{PC}$ has 4 generators. It is trivial to
generalize the result of group $O$ to group $O^{PC}.$

\section {Computational Procedure}
 Let us consider L-component operators which are classified into N types.
 Type means assembly
of operators with the same shape here( Any group element does not change the
shape of the operator). We'll restore all the bases of each
type into array A and  adopt the computational procedure as follows:

\begin{enumerate}
\item To produce a new L-component vector
standing for a
closed Wilson loop  which does not belong to A ( by comparing it
with the restored operators ).
\item To apply
96 elements of group $O^{PC}$ on this Wilson operator(named prototype)
to generate all the bases which belong to the same type and append these
bases to A. We also can get the representation matrix of these elements
in these bases.
\item
To construct new bases from the bases obtained from step 2 according
to PC=++, - +, + - and - -. Meanwhile, actually, we get a transform matrix
between old bases and new bases.  Therefore, we get the new representation
matrices which are diagonally respect to P and C by making a similar
transformation for the old ones obtained from step 2.
\item
For certain P and C, to construct new bases which are the combinations of
the bases obtained from step 3
in order to decompose the representation matrices into irreducible
representation $R^{PC}$.
We get the  multiplicity $m_{R^{PC}}$ of irreducible representation
$R^{PC}$ by our decompositing for
this representation. It is the same as
given by group theory:
\begin{flushright}
$ m_{R^{PC}}=\frac{1}{96}\sum\limits_K n_K \chi_K
\chi^{R^{PC}}_K,~~~~~~~~~~~~~~~~~~~~~~~~~~~~~~~~~~~~~~(3) $
\end{flushright}
where $K$ stands for the class, $n_K$ is the element number of $K^{th}$
class, $\chi_K$ is the character of $K^{th}$ class in this representation,
and $\chi^{R^{PC}}_K$ is the character of $K^{th}$ class
in the irreducible representation $R^{PC}$(see table 3);
\item
Return to step 1. We can sweep over all the L-component Wilson loops,
repeat 4 steps for new type, find
out all the types of operators  and
finish our procedure.
\end{enumerate}

We introduce steps 2, 3 and 4 in some detail here.

\vspace*{0.4cm}
(a) Step 2. To find the dimensionality and the bases for certain
prototype.

We illuminate it by the cubic group. The group has two  generators, which
means that all
elements $H\in O$ can be written as a product of some powers
of two generators,
$$H=C^{n_1}_4 C^{{n^\prime}_1}_3C^{n_2}_4 C^{{n^\prime}_2}_3\cdots
C^{n_m}_4 C^{{n^\prime}_m}_3 \cdots,$$
where $n_i=1,2,3,4$ and ${n^\prime}_i=1,2,3.$

Applying all possible products of $C_4$ and $C_3$ on one prototype, we
can find out complete operators in this type. Actually, 24 elements of
cubic group produce 24 operators. But, some of them are exactly the
same. All these operators have the same shape.
They construct the bases of
the representation in this type. Therefore, we can extract its
dimensionality and the representation  matrix of these elements in these
bases. The frame of the computer program is described in Fig.1.
\vspace*{0.4cm}
(b) Step 3. Assuming  the operation of
space parity P and charge conjugate C
on the vector are: $P\vv{a}=\vv{b},C\vv{d}=\vv{e}$,
therefore, we can construct the parity eigenvector: $\vv{a}+\vv{b}$ for
$P=+$
and $\vv{a}-\vv{b}$ for $P=-$, and $C$ conjugate eigenvector: $\vv{d}+\vv{e}$
for $C=+$ and $\vv{d}-\vv{e}$ for $C=-$.

Since $P$ and $C$ are commutative, one can find common eigenvector for $P$
and $C$. In this way, we may classify the bases according to $P$ and $C$
which are
the combinations of the bases obtained from step 2.

\vspace*{0.4cm}
(c) Step 4. With definite $P$ and $C$, we reduce the representation into
irreducible representation $R^{PC}$ of the $O^{PC}$ group
by solving zero eigenvalue of matrix equation.

We exemplify it by decomposing reducible representation into $T_1^{PC}$
representation (3 dimensions).

Denoting the reducible representation matrices of elements $C_4,~C_3$ just
by $C_4,~C_3$
themselves, we want to seek three vectors $\vv{a},\vv{b},\vv{c}$ to
construct the bases in which the representation matrices of
$C_3,~C_4$ are given by
Eq. (2d). Due to Eq. (2d), we have ( we assume that the
vector is
a row when it is left to the matrix and it is a column when right to the
matrix)
$$
 C_4\vv{a}=-\vv{c},~~~~~C_4\vv{b}=\vv{b},~~~~~C_4\vv{c}=\vv{a}.
$$
That is
\begin{flushright}
$
\begin{array}{lr}
\vv{a}C^T_4=-\vv{c},~~~~\vv{b}C^T_4=\vv{b},~~~~\vv{c}C^T_4=\vv{a}, &
~~~~~~~~~~~~~~~~~~(4)
\end{array}
$
\end{flushright}
here $'T'$ stands for the transpose of the matrix.

We can deduce out
\begin{flushright}
$
\begin{array}{lr}
(\vv{a}~\vv{b}~\vv{c})
    \left(
    \begin{array}{ccc}
        C^T_4~~  &  0  &  0\\
              0  &~~ C^T_4~~ & 0\\
              0  &  0   & ~~C^T_4
         \end{array}
     \right)
     =(\vv{a}~\vv{b}~\vv{c})
     \left(
        \begin{array}{ccc}
              ~~0 &  0  &  1\\
              0   & ~~1~~ & 0\\
              -1  &  0  & ~~0
        \end{array}
     \right)
     ,~~~~~~(5a)
\end{array}
$
\end{flushright}
or
\begin{flushright}
$ \begin{array}{lr}
     (\vv{a}~\vv{b}~\vv{c})
     \left(
         \begin{array}{ccc}
              C^T_4~~ & 0  &  -1\\
              0   & ~~C^T_4-1~~ & 0\\
              1   &  0 & ~~C^T_4
        \end{array}
     \right)
     =0
     .~~~~~~~~~~~~~~~~~~~~~~~~(5b)
  \end{array}
$
\end{flushright}
For $C_3$, we have analogously
$$
     (\vv{a}~\vv{b}~\vv{c})
     \left(
        \begin{array}{ccc}
              C^T_3~~ & 1  &  0\\
              0   & ~~C^T_3~~ & -1\\
              1   &   0  & ~~C^T_3
       \end{array}
     \right)
     =0.
$$
So
\begin{flushright}
$
\begin{array}{lr}
     (\vv{a}~\vv{b}~\vv{c})
     \left(
         \begin{array}{lccccr}
              C^T_4~~ &  0  &  -1  & ~~C^T_3~~  &  1  &  0\\
              0   & ~~C^T_4-1~~ & 0  &  0  & ~~C^T_3~~ &  -1\\
              1   &  0  & ~~C^T_4~~ &  1  &  0  & ~~C^T_3
        \end{array}
     \right)
     =0.
     ~~~~~~~~~~~(6)
\end{array}
$
\end{flushright}
Thus we can solve it with mathematical software. After getting the solution
$\vv{a},~\vv{b},~\vv{c}$ from eq. (6),  we find out the bases
of  representation $T_1^{PC}$ which are  our expected operators.

We emphasize that above idea is suitable for all the finite groups.

\section {Results}

We find the numbers of types of 4-,6-,8-,10-link operators are
1, 3, 18, 132, respectively. Table 1 gives the multiplicities of
the reducible representation corresponding to definite type for 4-, 6-, 8- and
10-link operators ( these reducible representations with the same
dimension
are not equivalent each other except for the 96-dimensional
representations).
\begin{center}
\begin{tabular}{|c|c|c|c|c|c|c|}\hline
dimension of the reducible rep. & ~6 & ~8 & ~12 & ~24 & ~48 & ~96 \\ \hline

 4-link  & 1   &  0  &  0    &   0   &   0   &  0      \\ \hline
6-link  & 0   &  1  &  1    &   1   &   0   &  0      \\ \hline
8-link  & 2   &  0  &  5    &   5    &   4    &   2   \\ \hline
10-link & 0   &  0  &  2   &   22    &   43    &   65 \\ \hline
\end{tabular}\\

Table 1  The multiplicities of the reducible representations for 4-, 6-, 8-
and 10-link operators.
\end{center}

As an example, we only show the results of one type of 10-link
operators here, the prototype of which is (1,2,-1,3,3,1,-2,-1,-3,-3). Its
dimensionality is 24.

From step 2, we get bases $\{ \vv{o_i}\}(i=1,\cdots,24)$ which belongs to
the same type as follows:\\
  1)  1 2 -1 3 3 1 -2 -1 -3 -3  ~   2)  -2 -1 2 -3 -3 -2 1 2 3 3
  3)  2 1 -2 3 3 2 -1 -2 -3 -3 \\    4)  -3 -3 -2 -1 2 3 3 -2 1 2
  5)  3 3 2 1 -2 -3 -3 2 -1 -2     ~6)  1 3 -1 -2 -2 1 -3 -1 2 2\\
  7)  -1 -3 1 2 2 -1 3 1 -2 -2     ~8)  -2 -2 1 3 -1 2 2 1 -3 -1
  ~9)  2 2 -1 -3 1 -2 -2 -1 3 1\\     10)  -3 2 3 1 1 -3 -2 3 -1 -1
  ~11)  3 -2 -3 -1 -1 3 2 -3 1 1   ~  12)  1 1 -3 2 3 -1 -1 -3 -2 3\\
 13)  -1 -1 3 -2 -3 1 1 3 2 -3    ~ 14)  3 -1 -3 -2 -2 3 1 -3 2 2
 ~15)  -3 1 3 2 2 -3 -1 3 -2 -2\\     16)  -2 -2 3 -1 -3 2 2 3 1 -3
 ~17)  2 2 -3 1 3 -2 -2 -3 -1 3   ~18)  2 3 -2 1 1 2 -3 -2 -1 -1\\
 19)  -2 -3 2 -1 -1 -2 3 2 1 1    ~ 20)  1 1 2 3 -2 -1 -1 2 -3 -2
 ~21)  -1 -1 -2 -3 2 1 1 -2 3 2\\     22)  -1 2 1 -3 -3 -1 -2 1 3 3
 ~23)  1 -2 -1 3 3 1 2 -1 -3 -3     ~24)  -3 -3 -1 2 1 3 3 -1 -2 1.

From step 3, new bases $\{ \vv{a_i} \}(i=1,2,\cdots,24)$ decomposed by
$P,~C$
are listed as follows. They are linear combinations of the old bases
$\vv{o_i}$. For
instance, we notify
 $\vv{a_{13}}=\vv{o_1}-\vv{o_{23}}-\vv{o_{22}}+\vv{o_{24}}$
 by (1,-23,-22,24). \\
1) 1  23  22  24~~         2)  2   4   3   5   ~~ 3) 6   8   7   9  ~~
4) 10  12  11  13\\    5) 14  16  15  17   ~~      6) 18  20 19  21 ~~
7) 18  20 -19 -21   ~~      8)  14  16 -15 -17\\    9) 10 12 -11 -13 ~~
10)  6   8  -7  -9   ~~11) 2   4  -3  -5     ~~ 12) 1  23 -22 -24\\
13)  1 -23 -22  24       ~~ 14) 2  -4   3 -5   ~~15) 6  -8   7  -9~~
16) 10 -12  11 -13\\   17) 14 -16 15 -17       ~~ 18) 18 -20  19 -21 ~~
19) 18 -20 -19  21   ~~     20) 14 -16 -15  17\\   21) 10 -12 -11  13 ~~
22) 6  -8  -7   9  ~~23) 2  -4  -3   5     ~~   24)  1 -23  22 -24.

From step 4, we get the irreducible operators which is the combination of
$\vv{a}_i$. the combination coefficients are shown in Table 2.
\begin{center}
{\footnotesize
\begin{tabular}{|c|c|c|c|c|}\hline
~~ & $PC=++$   &   $PC=-+$   &  $PC=+-$  &  $PC=--$ \\      &
$\{\vv{a}_i\},i=1,\cdots,6$ & $i=7,\cdots,12$ & $i=13,\cdots,18$  &
$i=19,\cdots,24$  \\ \hline
$A_1$  & $(1, 1, 1, 1, 1, 1)$  &   &     & $(1, 1, 1,1, 1, 1)$ \\ \hline
$A_2$  & $(1, -1, -1, -1, 1, 1)$  &   &  & $(1, 1, -1, -1, -1, 1)$\\ \hline
$E$ & $\left\{ \begin{array}{c} (-1, 1, 1, -2, 2, -1)\\(-3, -3, 3, 0, 0, 3)
\end{array}\right.$    &  &  &
$\left\{ \begin{array}{c}(2, -1, 1, -2, 1,
-1)\\(0, -3, -3, 0, 3, 3) \end{array}\right. $  \\    &
$\left\{
\begin{array}{c}(1, -1, 2, -1, 1, -2)\\(-3, -3, 0, 3, 3, 0)
\end{array}\right.$    &  &  &
$\left\{ \begin{array}{c} (1, 1, -1, -1, 2,
-2)\\(3, -3, -3, 3, 0, 0)\end{array}\right. $  \\ \hline
$T_1$ & & $\left\{
\begin{array}{c}(0, 0, 0, -1, 0, -1)\\(-1, 0, 0, 0,1,0)\\(0, -1, 1, 0, 0, 0)
\end{array}\right.$ & $\left\{ \begin{array}{c}(1, 0, 1, 0, 0, 0)\\(0, -1, 0,
0, 0, 1)\\(0, 0, 0, -1, 1, 0) \end{array}\right.$ & \\ \hline
$T_2$ & &
$\left\{ \begin{array}{c}(0,0,0,-1,0,1)\\(1,0,0,0,1,0)\\(0,1,1,0,0,0)
\end{array}\right.$ & ~$\left\{
\begin{array}{c}(1,0,-1,0,0,0)\\(0,1,0,0,0,1)\\(0, 0, 0, 1, 1, 0)
\end{array}\right.$ & \\ \hline
\end{tabular}\\
Table 2. Combination coefficients for the combination of $\{ \vv{a_i} \}$ }
\end{center}
We also present some results of 8-link operators in Appendix B. The
calculations for other operators are similar and interested reader can obtain
the complete results of 8-link, 10-link operators and computation program for
arbitrary-link operators from author D. Q. Liu.

\section {Conclusions}
Using our method, we can  combine  arbitrary-link operators
into irreducible representation of cubic group easily. If we accept the
opinion in Ref.[1], we have known the mapping between arbitrary-link
operators and the states with certain $J^{PC}(J\leq 3)$. In this point, we
have completed the combination of operators to states with certain
$J^{PC}(J\leq 3)$.

But, due to Wigner-Eckart theorem, the mapping is not one-to-one, even is
not definite because of the breaking of $SO(3)$ symmetry. The only we can say
is, for example, that there maybe exist some projection onto $2^{++}$ in
irreducible representation $E^{++}$ or $ T^{++}_2$. From our calculation, we
do not know the projection of which type of the operators onto a
definite $J^{PC}$ state are larger$^{\cite{s1}}$. One way to deal with
the problem
is the application of variational principle$^{[3]}$, or continue
limits$^{[4]}$. We will discuss it more detail in the next paper in which we
want to develop a approach named improved-operator one.


\newpage
\section *{Appendix A}
The cubic group $O$ has 24 elements which may be divided into five classes of
conjugate elements: $E,~C_2,~C_3,~C_4,~C^2_4$, and there are two generators
among the elements: $C_3$ and $C_4$. The group has five irreducible
representations, named $R=A_1,~A_2,~E,~T_1$ and $T_2$,
dimensions of which are 1, 1, 2, 3 and 3, respectively. $A_ 1$ is the
trivial and $T_1$ the vector representation in these representations. After
generalizing $O$ to $O^{PC}$, $O^{PC}$
 has 96 elements and 20 irreducible representation
$R^{PC}$(P and C have two
 probabilities:+ and -). For the simplification of
the discussion, we show
 character table for irreducible representation of the
cubic group $O$ here.

\begin{center}
\begin{tabular}{|c|c|c|c|c|c|}\hline
~~    &  $E$ & $6C_2$ & $8C_3$ & $6C_4$ & $3C^2_4$  \\ \hline
$A_1$ &  1   &   1     &  1     &   1    &   1      \\ \hline
$A_2$ &  1   &   -1    &   1    &  -1    &   1      \\ \hline
$E$   &  2   &   0     &   -1   &   0    &   2      \\ \hline
$T_1$ &  3   &   -1    &   0    &   1    &   -1     \\ \hline
$T_2$ &  3   &   1     &   0    &   -1   &   -1     \\ \hline
\end{tabular}\\
Table 3 The character table for the irreducible representation\\
 of the cubic group $O$.
\end{center}

\newpage
\section *{Appendix B}
We only present results for types of 8-link operators with 6-, 12- and
24-dimension representations here.\\

{\bf 1. operators with prototype (1 2 3 3 -2 -1 -3 -3).}\\
Bases $\{ \vv{o_i} \}$ are\\
     1)  1 2 3 3 -2 -1 -3 -3   ~~2)  2 1 3 3 -1 -2 -3 -3
   ~~3)  -3 -3 -2 -1 3 3 1 2   \\~~4)  3 3 2 1 -3 -3 -1 -2
   ~~5)  1 3 -2 -2 -3 -1 2 2 ~~ ~~6)  -1 -3 2 2 3 1 -2 -2\\
  ~~7)  -2 -2 1 3 2 2 -3 -1 ~~ ~~8)  2 2 -1 -3 -2 -2 3 1
   ~~9)  -3 2 1 1 -2 3 -1 -1\\   ~~10)  3 -2 -1 -1 2 -3 1 1
   ~~11)  1 1 -3 2 -1 -1 -2 3 ~~   12)  -1 -1 3 -2 1 1 2 -3\\
    13)  3 -1 -2 -2 1 -3 2 2  ~~  14)  -3 1 2 2 -1 3 -2 -2
 ~~  15)  -2 -2 3 -1 2 2 1 -3  \\   16)  2 2 -3 1 -2 -2 -1 3
   ~~17)  2 3 1 1 -3 -2 -1 -1  ~~  18)  -2 -3 -1 -1 3 2 1 1\\
    19)  1 1 2 3 -1 -1 -3 -2 ~~   20)  -1 -1 -2 -3 1 1 3 2
  ~~ 21)  -1 2 -3 -3 -2 1 3 3 \\   22)  1 -2 3 3 2 -1 -3 -3
  ~~ 23)  -3 -3 -1 2 3 3 -2 1 ~~   24)  3 3 1 -2 -3 -3 2 -1.\\
Bases $\{ \vv{a_i} \} (i=1,\cdots,24)$ are  \\
   1)     1   3   2   4        ~~2)      5   7   6   8
  ~~3)     9  11  10  12        ~~4)     13  15  14  16\\
 ~~5)    17  19  18  20       ~~6)     21  23  22  24
 ~~7)    21  23 -22 -24       ~~8)     17  19 -18 -20\\
 ~~9)    13  15 -14 -16      ~~10)      9  11 -10 -12
~~11)     5   7  -6  -8      ~~12)      1   3  -2  -4\\
~~13)     1  -3   2  -4      ~~14)      5  -7   6  -8
~~15)     9 -11  10 -12      ~~16)     13 -15  14 -16\\
~~17)    17 -19  18 -20      ~~18)     21 -23  22 -24
~~19)    21 -23 -22  24      ~~20)     17 -19 -18  20\\
~~21)    13 -15 -14  16      ~~22)      9 -11 -10  12
~~23)     5  -7  -6   8      ~~24)      1  -3  -2   4\\
Combination coefficients are presented in table 4.\\

{\bf 2. operators with prototype (1 2 -1 3 -1 -2 1 -3).}\\
Bases $\{ \vv{o_i} \}$ are\\
    1)  1 2 -1 3 -1 -2 1 -3    ~~2)  -2 -1 2 -3 2 1 -2 3
  ~~3)  -3 2 -1 -2 3 -2 1 2 \\  ~~4)  1 3 -1 -2 -1 -3 1 2
  ~~5)  -2 -1 3 1 2 1 -3 -1   ~~6)  -3 2 3 1 3 -2 -3 -1\\
  ~~7)  1 3 2 -3 -1 -3 -2 3   ~~8)  3 -1 -3 -2 -3 1 3 2
  ~~9)  -2 -3 -1 3 2 3 1 -3 \\  ~~10)  2 3 -2 1 -2 -3 2 -1
  ~~11)  1 -2 3 2 -1 2 -3 -2   ~~12)  -1 2 1 -3 1 -2 -1 3\\
  ~~13)  1 -3 -1 2 -1 3 1 -2   ~~14)  2 -1 -3 1 -2 1 3 -1
  ~~15)  -3 -2 3 -1 3 2 -3 1  \\ ~~16)  -1 3 -2 -3 1 -3 2 3
  ~~17)  -2 1 2 3 2 -1 -2 -3   ~~18)  3 2 1 -2 -3 -2 -1 2\\
  ~~19)  1 -2 -1 -3 -1 2 1 3   ~~20)  -3 -1 -2 1 3 1 2 -1
  ~~21)  -3 1 3 -2 3 -1 -3 2\\   ~~22)  -2 3 1 -3 2 -3 -1 3
  ~~23)  -2 -3 2 1 2 3 -2 -1   ~~24)  1 2 -3 -2 -1 -2 3 2.\\
Bases $\{ \vv{a_i} \}(i=1,\cdots,24)$ are  \\
   1)     1  12      ~~2)      2   3
 ~~3)     4   5     ~~4)      6   7
 ~~5)     8   9     ~~6)     10  11 \\
 ~~7)    13  14     ~~8)     15  16
 ~~9)    17  18    ~~10)     19  20
~~11)    21  22    ~~12)     23  24  \\
~~13)     1 -12    ~~14)      2  -3
~~15)     4  -5    ~~16)      6  -7
~~17)     8  -9    ~~18)     10 -11 \\
~~19)    13 -14    ~~20)     15 -16
~~21)    17 -18    ~~22)     19 -20
~~23)    21 -22    ~~24)     23 -24  \\
Combination coefficients are shown in table 5.\\

{\bf 3. Operators with prototype (1 2 3 3 -1 -2 -3 -3)}.\\
Bases $\{ \vv{o_i} \}$ are\\
    1)  1 2 3 3 -1 -2 -3 -3    ~~2)  -2 -1 -3 -3 2 1 3 3
  ~~3)  -3 -3 -1 -2 3 3 1 2 \\  ~~4)  1 3 -2 -2 -1 -3 2 2
  ~~5)  -2 -2 3 1 2 2 -3 -1   ~~6)  -3 2 1 1 3 -2 -1 -1\\
  ~~7)  1 1 2 -3 -1 -1 -2 3   ~~8)  3 -1 -2 -2 -3 1 2 2
  ~~9)  -2 -2 -1 3 2 2 1 -3  \\~~10)  2 3 1 1 -2 -3 -1 -1
  ~~11)  1 1 3 2 -1 -1 -3 -2   ~~12)  -1 2 -3 -3 1 -2 3 3\\
  ~~13)  -3 -3 2 -1 3 3 -2 1   ~~14)  -3 -1 2 2 3 1 -2 -2
  ~~15)  2 2 -1 -3 -2 -2 1 3 \\  ~~16)  -2 3 -1 -1 2 -3 1 1
  ~~17)  -1 -1 3 -2 1 1 -3 2   ~~18)  3 3 2 1 -3 -3 -2 -1\\
  ~~19)  1 -3 2 2 -1 3 -2 -2   ~~20)  2 2 -3 1 -2 -2 3 -1
  ~~21)  -3 -2 -1 -1 3 2 1 1  \\ ~~22)  -1 -1 -2 -3 1 1 2 3
  ~~23)  -2 1 3 3 2 -1 -3 -3   ~~24)  3 3 1 -2 -3 -3 -1 2.\\
Bases $\{ \vv{a_i} \} (i=1,\cdots,24)$ are  \\
   1)     1  18   ~~2)      2   3
 ~~3)     4   5  ~~4)      6   7
 ~~5)     8   9  ~~6)     10  11  \\
 ~~7)    12  13  ~~8)     14  15
 ~~9)    16  17  ~~10)     19  20
~~11)    21  22    ~~12)     23  24 \\
~~13)     1 -18   ~~14)      2  -3
~~15)     4  -5    ~~16)      6  -7
~~17)     8  -9    ~~18)     10 -11  \\
~~19)    12 -13  ~~20)     14 -15
~~21)    16 -17   ~~22)     19 -20
~~23)    21 -22    ~~24)     23 -24  \\
Combination coefficients are given in table 6.\\

{\bf 4. Operators with prototype (1 2 -1 -2 1 3 -1 -3 ).}\\
Bases $\{ \vv{o_i} \}$ are\\
    1)  1 2 -1 -2 1 3 -1 -3     ~~2)  -2 -1 2 1 -2 -3 2 3
   ~~ 3)  2 1 -2 -1 2 3 -2 -3  \\   4)  -3 -2 3 2 -1 -2 1 2
    ~~5)  3 2 -3 -2 1 2 -1 -2  ~~   6)  1 3 -1 -3 1 -2 -1 2\\
    ~~7)  -1 -3 1 3 -1 2 1 -2 ~~    8)  -2 1 2 -1 3 1 -3 -1
    ~~9)  2 -1 -2 1 -3 -1 3 1   \\  10)  -3 2 3 -2 -3 1 3 -1
    ~~11)  3 -2 -3 2 3 -1 -3 1    ~~ 12)  1 -3 -1 3 2 -3 -2 3\\
    13)  -1 3 1 -3 -2 3 2 -3 ~~    14)  2 -1 -2 1 2 3 -2 -3
    ~~15)  -2 1 2 -1 -2 -3 2 3\\     16)  3 2 -3 -2 -1 2 1 -2
    ~~17)  -3 -2 3 2 1 -2 -1 2 ~~    18)  -1 3 1 -3 -1 2 1 -2\\
    19)  1 -3 -1 3 1 -2 -1 2   ~~  20)  -2 1 2 -1 -3 1 3 -1
    ~~21)  3 2 -3 -2 3 -1 -3 1 \\    22)  -3 -2 3 2 -3 1 3 -1
    ~~23)  -1 3 1 -3 2 3 -2 -3  ~~   24)  1 -3 -1 3 -2 -3 2 3.\\
Bases $\{ \vv{a_i} \}(i=1,\cdots,12)$ are  \\
   1)     1  18  19  20 ~~   2)      2   4   3   5
  ~~ 3)     6   8   7   9 ~~        4)     10  12  11  13\\
   5)    14  16  15  17  ~~       6)     21  23  22  24
   ~~7)    21  23 -22 -24~~         8)     14  16 -15 -17\\
   9)    10  12 -11 -13 ~~       10)      6   8  -7  -9
  ~~11)     2   4  -3  -5     ~~   12)      1  18 -19 -20\\
  13)     1 -18 -19  20  ~~      14)      2  -4   3  -5
  ~~15)     6  -8   7  -9   ~~     16)     10 -12  11 -13\\
  17)    14 -16  15 -17 ~~       18)     21 -23  22 -24
  ~~19)    21 -23 -22  24~~        20)     14 -16 -15  17\\
  21)    10 -12 -11  13   ~~     22)      6  -8  -7   9
  ~~23)     2  -4  -3   5 ~~       24)      1 -18  19 -20.\\
Combination coefficients are shown in table 7.\\

{\bf 5. Operators with prototype (1 2 -1 2 -1 -2 -2 1 ).}\\
Bases $\{ \vv{o_i} \}$ are\\
    1)  1 2 -1 2 -1 -2 -2 1  ~~   2)  -2 -1 2 -1 2 1 1 -2
    ~~3)  2 1 -2 1 -2 -1 -1 2    \\ 4)  2 -1 -1 -2 1 -2 1 2
    ~~5)  1 3 -1 3 -1 -3 -3 1 ~~    6)  -1 -3 1 -3 1 3 3 -1
    \\7)  -1 3 3 1 -3 1 -3 -1~~     8)  1 -3 -3 -1 3 -1 3 1
   ~~ 9)  -3 2 3 2 3 -2 -2 -3    \\ 10)  3 -2 -3 -2 -3 2 2 3
    ~~11)  3 2 2 -3 -2 -3 -2 3   ~~  12)  -3 -2 -2 3 2 3 2 -3
    \\13)  3 -1 -3 -1 -3 1 1 3~~     14)  -3 1 3 1 3 -1 -1 -3
    ~~15)  -3 -1 -1 3 1 3 1 -3    \\ 16)  3 1 1 -3 -1 -3 -1 3
    ~~17)  2 3 -2 3 -2 -3 -3 2~~     18)  -2 -3 2 -3 2 3 3 -2
    \\19)  -2 3 3 2 -3 2 -3 -2    ~~ 20)  2 -3 -3 -2 3 -2 3 2
    ~~21)  -1 2 1 2 1 -2 -2 -1\\     22)  1 -2 -1 -2 -1 2 2 1
    ~~23)  1 2 2 -1 -2 -1 -2 1    ~~ 24)  -1 -2 -2 1 2 1 2 -1.\\
Bases $\{ \vv{a_i} \}(i=1,\cdots,12)$ are  \\
   1)     1   3   2   4~~         2)      5   7   6   8
   ~~3)     9  11  10  12  ~~       4)     13  15  14  16
   \\5)    17  19  18  20~~         6)     21  23  22  24
   ~~7)    21  23 -22 -24    ~~     8)     17  19 -18 -20
   \\9)    13  15 -14 -16      ~~  10)      9  11 -10 -12
  ~~11)     5   7  -6  -8        ~~12)      1   3  -2  -4
  \\13)     1  -3  -2   4~~        14)      5  -7   6  -8
  ~~15)     9 -11  10 -12  ~~      16)     13 -15  14 -16
  \\17)    17 -19  18 -20    ~~    18)     21 -23  22 -24
  ~~19)    21 -23 -22  24      ~~  20)     17 -19 -18  20
  \\21)    13 -15 -14  16        ~~22)      9 -11 -10  12
  ~~23)     5  -7  -6   8        ~~24)      1  -3   2  -4\\
Combination coefficients are given in table 8.\\

{\bf 6. operators with prototype (1 2 -1 -2 -2 -1 2 1 ).}\\
Bases $\{ \vv{o_i} \}$ are\\
    1)  1 2 -1 -2 -2 -1 2 1    ~~2)  2 1 -2 -1 -1 -2 1 2
  ~~3)  1 3 -1 -3 -3 -1 3 1  \\ ~~4)  -1 -3 1 3 3 1 -3 -1
  ~~5)  -3 2 3 -2 -2 3 2 -3   ~~6)  3 -2 -3 2 2 -3 -2 3\\
  ~~7)  3 -1 -3 1 1 -3 -1 3   ~~8)  -3 1 3 -1 -1 3 1 -3
  ~~9)  2 3 -2 -3 -3 -2 3 2 \\  ~~10)  -2 -3 2 3 3 2 -3 -2
  ~~11)  -1 2 1 -2 -2 1 2 -1   ~~12)  1 -2 -1 2 2 -1 -2 1.\\
Bases $\{ \vv{a_i} \}(i=1,\cdots,12)$ are  \\
   1)     1   2      ~~2)      3   4
 ~~3)     5   6     ~~4)      7   8
 ~~5)     9  10     ~~6)     11  12  \\
 ~~7)    11 -12     ~~8)      9 -10
 ~~9)     7  -8    ~~10)      5  -6
~~11)     3  -4    ~~12)      1  -2   \\
Combination coefficients are shown in table 9.\\

{\bf 7. operators with prototype (1 2 -1 3 1 -2 -1 -3).}\\
Bases $\{ \vv{o_i} \}$ are\\
    1)  1 2 -1 3 1 -2 -1 -3   ~~  2)  -2 -1 2 -3 -2 1 2 3
    ~~3)  2 1 -2 3 2 -1 -2 -3     \\4)  -3 -2 -1 2 3 -2 1 2
    ~~5)  3 2 1 -2 -3 2 -1 -2~~     6)  1 3 -1 -2 1 -3 -1 2
    \\7)  -2 1 3 -1 2 1 -3 -1    ~~ 8)  2 -1 -3 1 -2 -1 3 1
    ~~9)  -3 2 3 1 -3 -2 3 -1\\     10)  3 -2 -3 -1 3 2 -3 1
    ~~11)  1 -3 2 3 -1 -3 -2 3   ~~  12)  -1 3 -2 -3 1 3 2 -3.\\
Bases $\{ \vv{a_i} \}(i=1,\cdots,12)$ are  \\
  1)     1   8   6   7~~         2)      2   4   3   5
 ~~  3)     9  11  10  12 ~~        4)      9  11 -10 -12
   \\5)     2   4  -3  -5   ~~      6)      1   8  -6  -7
   ~~7)     1  -8   6  -7     ~~    8)      2  -4   3  -5
   \\9)     9 -11  10 -12       ~~ 10)      9 -11 -10  12
  ~~11)     2  -4  -3   5        12)      1  -8  -6   7.\\
Combination coefficients are given in table 10.\\

{\bf 8. operators with prototype (1 2 -1 -1 -1 -2 1 1 ).}\\
Bases $\{ \vv{o_i} \}$ are\\
    1)  1 2 -1 -1 -1 -2 1 1 ~~    2)  -2 -1 2 2 2 1 -2 -2
    ~~3)  2 2 -1 -2 -2 -2 1 2   \\  4)  1 3 -1 -1 -1 -3 1 1
    ~~5)  -1 -1 3 1 1 1 -3 -1    ~~ 6)  -3 2 3 3 3 -2 -3 -3
    \\7)  3 3 2 -3 -3 -3 -2 3     ~~8)  3 -1 -3 -3 -3 1 3 3
    ~~9)  -3 -3 -1 3 3 3 1 -3     \\10)  2 3 -2 -2 -2 -3 2 2
    ~~11)  -2 -2 3 2 2 2 -3 -2~~     12)  -1 2 1 1 1 -2 -1 -1.\\
Bases $\{ \vv{a_i} \}(i=1,\cdots,12)$ are  \\
   1)     1  12 ~~2) 2   3 ~~3) 4   5 ~~4)      6   7
   ~~5)     8   9 ~~ 6)     10  11 \\
   ~~7)     1 -12 ~~8) 2  -3
   ~~9)     4  -5 ~~10) 6  -7~~
  11)     8  -9 12) 10 -11.\\
Combination coefficients are given in table 11.\\

{\bf 9. operators with prototype (1 2 -1 -2 -1 -2 1 2 ).}\\
Bases $\{ \vv{o_i} \}$ are\\
    1)  1 2 -1 -2 -1 -2 1 2    ~~ 2)  -2 -1 2 1 2 1 -2 -1~~
    3)  1 3 -1 -3 -1 -3 1 3\\     4)  -3 -1 3 1 3 1 -3 -1
    ~~5)  -3 2 3 -2 3 -2 -3 2  ~~   6)  -2 3 2 -3 2 -3 -2 3
    \\7)  3 -1 -3 1 -3 1 3 -1 ~~    8)  1 -3 -1 3 -1 3 1 -3
    ~~9)  2 3 -2 -3 -2 -3 2 3    \\ 10)  -3 -2 3 2 3 2 -3 -2
    ~~11)  -1 2 1 -2 1 -2 -1 2   ~~  12)  -2 1 2 -1 2 -1 -2 1.\\
Bases $\{ \vv{a_i} \}(i=1,\cdots,12)$ are  \\
   1)     1   2  ~~2)  3   4~~3)     5   6 ~~4) 7  8
~~5)     9  10   ~~6) 11  12  11  12
\\   7)     1  -2 ~~8)  3  -4  ~~ 9) 5  -6 ~~ 10)  7  -8
  ~~11)     9 -10 ~~12) 11 -12.\\
Combination coefficients are given in table 12.\\

{\bf 10. operators with prototype (1 2 -1 -2 1 -2 -1 2 ).}\\
Bases $\{ \vv{o_i} \}$ are\\
    1)  1 2 -1 -2 1 -2 -1 2 ~~    2)  -2 -1 2 1 -2 1 2 -1
    ~~3)  2 1 -2 -1 2 -1 -2 1   \\  4)  1 3 -1 -3 1 -3 -1 3
    ~~5)  -1 -3 1 3 -1 3 1 -3 ~~    6)  -3 2 3 -2 -3 -2 3 2
    \\7)  3 -2 -3 2 3 2 -3 -2     ~~8)  3 -1 -3 1 3 1 -3 -1
    ~~9)  -3 1 3 -1 -3 -1 3 1 \\    10)  2 3 -2 -3 2 -3 -2 3
    ~~11)  -2 -3 2 3 -2 3 2 -3  ~~   12)  -1 2 1 -2 -1 -2 1 2.\\
Bases $\{ \vv{a_i} \}(i=1,\cdots,12)$ are  \\
   1)     1  12   1  12  ~~       2)      2   3
~~   3)     4   5   4   5    ~~     4)      6   7
  ~~ 5)     8   9   8   9      ~~   6)     10  11
   \\7)    10 -11  10 -11        ~~ 8)      8  -9
~~   9)     6  -7   6  -7        ~~10)      4  -5
  ~~11)     2  -3   2  -3        ~~12) 1 -12.\\
Combination coefficients are given in table 13.\\

{\bf 11. operators with prototype (1 2 2 -1 -1 -2 -2 1).}\\
Bases $\{ \vv{o_i} \}$ are\\
 1)  1 2 2 -1 -1 -2 -2 1 ~~    2)  -2 -1 -1 2 2 1 1 -2
    ~~3)  1 3 3 -1 -1 -3 -3 1\\     4)  -1 3 3 1 1 -3 -3 -1
    ~~5)  -3 2 2 3 3 -2 -2 -3   ~~  6)  3 2 2 -3 -3 -2 -2 3.\\
Bases $\{ \vv{a_i} \}(i=1,\cdots,6)$ are  \\
  1)     1   2 ~~2)  3   4 ~~ 3)  5   6 ~~ 4) ~~ 1  -2
   ~~5)  3  -4  ~~ 6)      5  -6.\\
Combination coefficients are shown in table 14.\\

{\bf 12. operators with prototype (1 2 -1 -2 1 2 -1 -2 ).}\\
Bases $\{ \vv{o_i} \}$ are\\
    1)  1 2 -1 -2 1 2 -1 -2~~     2)  -2 -1 2 1 -2 -1 2 1
  ~~  3)  1 3 -1 -3 1 3 -1 -3  \\   4)  3 1 -3 -1 3 1 -3 -1
~~    5)  -3 2 3 -2 -3 2 3 -2    ~~ 6)  2 -3 -2 3 2 -3 -2 3.\\
Bases $\{ \vv{a_i} \}(i=1,\cdots,6)$ are  \\
   1)     1   2 ~~2)      3   4
  ~~ 3)     5   6 ~~4)      1  -2
   ~~5)     3  -4 ~~6)      5  -6.\\
Combination coefficients are shown in table 15.\\

\newpage
\begin{center}
{\scriptsize
\begin{tabular}{|c|c|c|c|c|}\hline
~~~~ &  $PC=++$   &   $PC=-+$   &  $PC=+-$  &  $PC=--$ \\
     & $\{\vv{a}_i\},i=1,\cdots,6$ & $i=7,\cdots,12$ & $i=13,\cdots,18$  &
$i=19,\cdots,24$  \\ \hline $A_1$ & $(1, 1, 1, 1, 1, 1)$ & & & $(1, 1, 1, 1,
1, 1)$  \\ \hline
$E$  &
 $\left\{ \begin{array}{c} (0, -1, 1, -1, 1, 0)\\
(2, -1, -1, -1, -1, 2) \end{array}\right.$ &&&
 $\left\{ \begin{array}{c} (0, 1, -1, 1, -1, 0)\\
(2, -1, -1, -1, -1, 2) \end{array}\right.$
 \\ \hline
$T_1$ & &
 $\left\{ \begin{array}{c} (1, 0, -1, 0, -1, -1)\\
(1, -1, 0, 1, 0, 1)\\(0, 1, -1, 1, 1, 0) \end{array}\right.$ &
 $\left\{ \begin{array}{c} (-1, -1, 0, -1, 0, 1)\\
(1, 0, 1, 0, -1, 1)\\(0, 1, 1, -1, 1, 0) \end{array}\right.$ & ~~
 \\ \hline
$T_2$ &
 $\left\{ \begin{array}{c} (0, 0, 1, 0, -1, 0)\\
(0, -1, 0, 1, 0, 0)\\(-1, 0, 0, 0, 0, 1) \end{array}\right.$ &
$\left\{ \begin{array}{c} (1, 0, 1, 0, 1, -1)\\
(-1, -1, 0, 1, 0, -1)\\(0, -1, -1, -1, 1, 0) \end{array}\right.$ &
 $\left\{ \begin{array}{c}(1, -1, 0, -1, 0, -1)\\
(1, 0, -1, 0, 1, 1)\\(0, -1, 1, 1, 1, 0)  \end{array}\right.$ &
 $\left\{ \begin{array}{c}(0, 1, 0, -1, 0, 0)\\
(0, 0, -1, 0, 1, 0)\\(-1, 0, 0, 0, 0, 1)  \end{array}\right.$
 \\ \hline
\end{tabular}} \\
Table 4 Combination coefficients for the combination\\
 of $\{\vv{a_i}\}$ from type 1.\\
\vspace*{0.4cm}

{\scriptsize
\begin{tabular}{|c|c|c|}\hline
~~~~ &  $PC=++$   &     $PC=+-$  \\
     & $\{\vv{a}_i\},i=1,\cdots,12$  & $i=13,\cdots,24$   \\ \hline
$A_1$ &
 $(1, 1, 1, 1, 1, 1, 1, 1, 1, 1, 1, 1)$ &~
 \\ \hline
$A_2$ &
$(1, -1, -1, -1, 1, 1, -1, -1, -1, 1, 1, 1)$ &~
 \\ \hline
$E$ &
 $ \begin{array}{c} \left\{ \begin{array}{c}
(-1, 1, 1, -2, 2, -1, 1, -2, 1, -1, 2, -1)\\
(-3, -3, 3, 0, 0, 3, 3, 0, -3, -3, 0, 3) \end{array}\right. \\
 \left\{ \begin{array}{c} (1, -1, 2, -1, 1, -2, 2, -1, -1, 1, 1, -2)\\
(-3, -3, 0, 3, 3, 0, 0, 3, -3, -3, 3, 0) \end{array}\right.\end{array} $ &~
 \\ \hline
$T_1$ &
 $ \left\{ \begin{array}{c} (0, 1, 0, 0, 1, 0, 0, 0, -1, 0, -1, 0)\\
(-1, 0, 0, -1, 0, 0, 0, 1, 0, 1, 0, 0)\\
(0, 0, -1, 0, 0, -1, 1, 0, 0, 0, 0, 1) \end{array}\right. $ &
$ \begin{array}{c} \left\{ \begin{array}{c}
(-1, 0, -1, 0, 0, 0, -1, 0, 0, -1, 0, 0)\\
(0, 1, 0, 0, 0, -1, 0, 0, 1, 0, 0, 1)\\
(0, 0, 0, 1, -1, 0, 0, 1, 0, 0, 1, 0)  \end{array}\right. \\
 \left\{ \begin{array}{c} (0, 0, 0, -1, 0, -1, 0, 1, 0, 0, 0, -1)\\
(0, 0, 1, 0, 1, 0, -1, 0, 0, 0, 1, 0)\\
(-1, 1, 0, 0, 0, 0, 0, 0, -1, 1, 0, 0) \end{array}\right. \end{array}$
 \\ \hline
$T_2$ &
 $ \left\{ \begin{array}{c} (0, -1, 0, 0, 1, 0, 0, 0, 1, 0, -1, 0)\\
(-1, 0, 0, 1, 0, 0, 0, -1, 0, 1, 0, 0)\\
(0, 0, 1, 0, 0, -1, -1, 0, 0, 0, 0, 1) \end{array}\right. $ &
$\begin{array}{c} \left\{ \begin{array}{c}
(-1, 0, 1, 0, 0, 0, 1, 0, 0, -1, 0, 0)\\
(0, -1, 0, 0, 0, -1, 0, 0, -1, 0, 0, 1)\\
(0, 0, 0, -1, -1, 0, 0, -1, 0, 0, 1, 0)\end{array}\right. \\
  \left\{ \begin{array}{c} (0, 0, 0, 1, 0, -1, 0, -1, 0, 0, 0, -1)\\
(0, 0, -1, 0, 1, 0, 1, 0, 0, 0, 1, 0)\\
(-1, -1, 0, 0, 0, 0, 0, 0, 1, 1, 0, 0) \end{array}\right.\end{array} $ \\ \hline
 \end{tabular}
 }  \\
Table 5 Combination coefficients for the combination\\
 of $\{ \vv{a_i}\}$ of type 2.\\
\vspace*{0.4cm}

{\scriptsize
\begin{tabular}{|c|c|c|}\hline
~~~~ &  $PC=++$   &     $PC=+-$  \\
     & $\{\vv{a}_i\},i=1,\cdots,12$  & $i=13,\cdots,24$   \\ \hline
$A_1$ & $(1, 1, 1, 1, 1, 1, 1, 1, 1, 1, 1, 1)$ & ~~
\\ \hline
$A_2$ & &
$(-1, 1, 1, 1, -1, -1, -1, -1, -1, 1, 1, 1)$
\\ \hline
$E$ &
 $ \left\{ \begin{array}{c} (0, 0, -1, 1, -1, 1, 0, -1, 1, -1, 1, 0)\\
(2, 2, -1, -1, -1, -1, 2, -1, -1, -1, -1, 2) \end{array}\right. $ &
 $  \left\{ \begin{array}{c}(2, -2, 1, 1, -1, -1, 2, -1, -1, 1, 1, -2)\\
(0, 0, -3, 3, 3, -3, 0, 3, -3, -3, 3, 0) \end{array}\right.$
\\ \hline
$T_1$ &
 $ \left\{ \begin{array}{c} (-1, 1, -1, 0, 1, 0, 1, 1, 0, -1, 0, -1)\\
(-1, 1, 0, -1, 0, -1, -1, 0, 1, 0, 1, 1)\\
(0, 0, -1, 1, -1, -1, 0, 1, -1, 1, 1, 0) \end{array}\right. $ &
$\begin{array}{c} \left\{ \begin{array}{c}
(0, 0, 0, 1, 0, 1, 0, 0, -1, 0, -1, 0)\\
(0, 0, -1, 0, -1, 0, 0, 1, 0, 1, 0, 0)\\
(1, -1, 0, 0, 0, 0, -1, 0, 0, 0, 0, 1)\end{array}\right. \\
  \left\{ \begin{array}{c} (-1, -1, -1, 0, -1, 0, 1, -1, 0, -1, 0, 1)\\
(1, 1, 0, 1, 0, -1, 1, 0, 1, 0, -1, 1)\\
(0, 0, 1, 1, -1, 1, 0, 1, 1, -1, 1, 0)\end{array}\right.\end{array} $
\\ \hline
$T_2$ &
$\begin{array}{c}
\left\{ \begin{array}{c} (0, 0, 0, 1, 0, -1, 0, 0, 1, 0, -1, 0)\\(0, 0, -1,
0, 1, 0, 0, -1, 0, 1, 0, 0)\\ (-1, -1, 0, 0, 0, 0, 1, 0, 0, 0, 0,
1)\end{array}
\right. \\   \left\{ \begin{array}{c} (1, -1, -1, 0, 1, 0, -1, 1,
0, -1, 0, 1)\\ (-1, 1, 0, 1, 0, 1, -1, 0, -1, 0, -1, 1)\\
(0, 0, 1, 1, 1, -1, 0, -1, -1, -1, 1, 0)\end{array}\right.\end{array} $ &
 $ \left\{ \begin{array}{c} (1, 1, -1, 0, -1, 0, -1, -1, 0, -1, 0, -1)\\
(1, 1, 0, -1, 0, 1, 1, 0, -1, 0, 1, 1)\\
(0, 0, -1, 1, 1, 1, 0, -1, 1, 1, 1, 0) \end{array}\right. $
 \\ \hline
 \end{tabular}
} \\
Table 6 Combination coefficients for the combination\\
 of $\{ \vv{a_i} \}$ from type 3.\\
\vspace*{0.4cm}

{\scriptsize
\begin{tabular}{|c|c|c|c|c|}\hline
~~~~ &  $PC=++$   &   $PC=-+$   &  $PC=+-$  &  $PC=--$ \\
     & $\{\vv{a}_i\},i=1,\cdots,6$ & $i=7,\cdots,12$ & $i=13,\cdots,18$  &
$i=19,\cdots,24$  \\ \hline
$A_1$ & $(1, 1, 1, 1, 1, 1)$ & & &  \\ \hline
$A_2$ & & & & $(-1, -1, -1, -1, -1, 1)$ \\ \hline
$E$  &
 $\left\{ \begin{array}{c}(1, -1, 1, 0, -1, 0)\\
(-1, -1, -1, 2, -1, 2) \end{array}\right.$ &~ &~ &
 $\left\{ \begin{array}{c} (2, -1, 2, -1, -1, 1)\\
(0, 3, 0, -3, 3, 3) \end{array}\right.$
 \\ \hline
$T_1$ & ~&
 $\left\{ \begin{array}{c} (-1, -1, 1, 0, -1, 0)\\
(1, 0, 1, -1, 0, 1)\\
(0, 1, 0, 1, -1, 1) \end{array}\right.$ &
 $\left\{ \begin{array}{c} (0, 0, 0, -1, 0, -1)\\
(0, -1, 0, 0, 1, 0)\\
(-1, 0, 1, 0, 0, 0) \end{array}\right.$ & ~~
 $\left\{ \begin{array}{c} (0, 0, 0, -1, 0, -1)\\
(0, -1, 0, 0, 1, 0)\\
(-1, 0, 1, 0, 0, 0) \end{array}\right.$
\\ \hline
$T_2$ &
 $\left\{ \begin{array}{c} (-1, 0, 1, 0, 0, 0)\\
(0, -1, 0, 0, 1, 0)\\(0, 0, 0, -1, 0, 1) \end{array}\right.$ &
$\left\{ \begin{array}{c} (1, -1, -1, 0, -1, 0)\\
(1, 0, 1, 1, 0, -1)\\
(0, -1, 0, 1, 1, 1) \end{array}\right.$ &
 $\left\{ \begin{array}{c}(0, -1, 0, 1, -1, -1)\\
(-1, 0, -1, 1, 0, 1)\\
(-1, -1, 1, 0, 1, 0)  \end{array}\right.$ &
 \\ \hline
\end{tabular}} \\
Table 7 Combination coefficients for the combination\\
 of $\{\vv{a_i}\}$ from type 4.\\
\vspace*{0.4cm}

{\scriptsize
\begin{tabular}{|c|c|c|c|c|}\hline
~~~~ &  $PC=++$   &   $PC=-+$   &  $PC=+-$  &  $PC=--$ \\
     & $\{\vv{a}_i\},i=1,\cdots,6$ & $i=7,\cdots,12$ & $i=13,\cdots,18$  &
$i=19,\cdots,24$  \\ \hline
$A_1$ & $(1, 1, 1, 1, 1, 1)$ & ~& ~&  \\ \hline
$A_2$ & ~& ~& $(1, -1, -1, 1, 1, 1)$ & \\ \hline
$E$  &
 $\left\{ \begin{array}{c}(0, -1, 1, -1, 1, 0)\\
(2, -1, -1, -1, -1, 2)
 \end{array}\right.$ &~ &
 $\left\{ \begin{array}{c}(-2, -1, -1, 1, 1, -2)\\
(0, 3, -3, -3, 3, 0) \end{array}\right.$ &
 \\ \hline
$T_1$ & ~&
 $\left\{ \begin{array}{c} (-1, 0, -1, 0, 1, 1)\\
(1, 1, 0, 1, 0, 1)\\
(0, 1, 1, -1, 1, 0) \end{array}\right.$ &
 $\left\{ \begin{array}{c} (0, 0, -1, 0, -1, 0)\\
(0, 1, 0, 1, 0, 0)\\
(-1, 0, 0, 0, 0, 1) \end{array}\right.$ & ~~
 $\left\{ \begin{array}{c} (1, 0, -1, 0, -1, -1)\\
(1, -1, 0, 1, 0, 1)\\
(0, 1, -1, 1, 1, 0) \end{array}\right.$
\\ \hline
$T_2$ &
 $\left\{ \begin{array}{c}(0, 0, 1, 0, -1, 0)\\
(0, -1, 0, 1, 0, 0)\\
(-1, 0, 0, 0, 0, 1)
\end{array}\right.$ &
$\left\{ \begin{array}{c} (-1, 0, 1, 0, -1, 1)\\
(-1, 1, 0, 1, 0, -1)\\
(0, -1, 1, 1, 1, 0) \end{array}\right.$ & &
 $\left\{ \begin{array}{c}(1, 0, 1, 0, 1, -1)\\
(-1, -1, 0, 1, 0, -1)\\
(0, -1, -1, -1, 1, 0)  \end{array}\right.$
 \\ \hline
\end{tabular}} \\
Table 8 Combination coefficients for the combination\\
 of $\{\vv{a_i}\}$ from type 5.\\
\vspace*{0.4cm}

{\footnotesize
\begin{tabular}{|c|c|c|}\hline
~~~~ &  $PC=++$   &     $PC=+-$  \\
     & $\{\vv{a}_i\},i=1,\cdots,6$  & $i=7,\cdots,12$   \\ \hline
$A_1$  &
$(1, 1, 1, 1, 1, 1)$ &
\\ \hline
$E$   &
 $ \left\{ \begin{array}{c} (0, -1, 1, -1, 1, 0)\\
(2, -1, -1, -1, -1, 2) \end{array}\right. $ &
\\ \hline
$T_1$  & &
 $ \left\{ \begin{array}{c} (1, 0, -1, 0, -1, -1)\\
(1, -1, 0, 1, 0, 1)\\(0, 1, -1, 1, 1, 0) \end{array}\right. $
\\ \hline
$T_2$  &
 $ \left\{ \begin{array}{c}(0, 0, 1, 0, -1, 0)\\
(0, -1, 0, 1, 0, 0)\\(-1, 0, 0, 0, 0, 1)  \end{array}\right. $ &
 $ \left\{ \begin{array}{c}(1, 0, 1, 0, 1, -1)\\
(-1, -1, 0, 1, 0, -1)\\(0, -1, -1, -1, 1, 0)  \end{array}\right. $
\\ \hline
\end{tabular}
}\\
Table 9 Combination coefficients for the combination\\
 of $\{ \vv{a_i} \}$ from type 6.\\
\vspace*{0.4cm}

{\scriptsize
\begin{tabular}{|c|c|c|c|c|}\hline
~~~~ &  $PC=++$   &   $PC=-+$   &  $PC=+-$  &  $PC=--$ \\
     & $\{\vv{a}_i\},i=1,\cdots,3$ & $i=4,\cdots,6$ & $i=7,\cdots,9$  &
$i=10,\cdots,12$  \\ \hline
$A_1$ & $(1, 1, 1)$ & ~& ~&  \\ \hline
$A_2$ & ~& ~&~& $(-1, -1, 1)$  \\ \hline
$E$  &
 $\left\{ \begin{array}{c}(1, -1, 0)\\
(-1, -1, 2)
 \end{array}\right.$ &~ & &
 $\left\{ \begin{array}{c}(2, -1, 1)\\(0, 3, 3) \end{array}\right.$
 \\ \hline
$T_1$ & ~&~&
 $\left\{ \begin{array}{c} (-1, 0, 0)\\(0, 1, 0)\\(0, 0, 1)
\end{array}\right.$ &
\\ \hline
$T_2$ & ~&
 $\left\{ \begin{array}{c}(0, 0, 1)\\(0, 1, 0)\\(1, 0, 0)
\end{array}\right.$ & &
 \\ \hline
\end{tabular}} \\
Table 10 Combination coefficients for the combination\\
 of $\{\vv{a_i}\}$ from type 7.\\
\vspace*{0.4cm}

{\footnotesize
\begin{tabular}{|c|c|c|}\hline
~~~~ &  $PC=++$   &     $PC=+-$  \\
     & $\{\vv{a}_i\},i=1,\cdots,6$  & $i=7,\cdots,12$   \\ \hline
$A_1$  &
$(1, 1, 1, 1, 1, 1)$ &
\\ \hline
$A_2$ &
$(1, -1, -1, -1, 1, 1)$
& \\ \hline
$E$   &
 $\begin{array}{c}
\left\{ \begin{array}{c}(-1, 1, 1, -2, 2, -1)\\
(-3, -3, 3, 0, 0, 3)\end{array}\right. \\
\left\{ \begin{array}{c}(1, -1, 2, -1, 1, -2)\\
(-3, -3, 0, 3, 3, 0)\end{array}\right. \end{array}$
&
\\ \hline
$T_1$  & &
 $ \left\{ \begin{array}{c} (0, 0, 0, -1, 0, -1)\\
(0, 0, 1, 0, 1, 0)\\
(-1, 1, 0, 0, 0, 0)\end{array}\right. $
\\ \hline
$T_2$  & &
 $ \left\{ \begin{array}{c}(0, 0, 0, -1, 0, 1)\\
(0, 0, 1, 0, -1, 0)\\
(1, 1, 0, 0, 0, 0)\end{array}\right. $
\\ \hline
\end{tabular}
}\\
Table 11 Combination coefficients for the combination\\
 of $\{ \vv{a_i} \}$ from type 8.\\
\vspace*{0.4cm}

{\footnotesize
\begin{tabular}{|c|c|c|}\hline
~~~~ &  $PC=++$   &     $PC=+-$  \\
     & $\{\vv{a}_i\},i=1,\cdots,6$  & $i=7,\cdots,12$   \\ \hline
$A_1$  &
$(1, 1, 1, 1, 1, 1)$ &
\\ \hline
$A_2$ &~&
$(1, -1, -1, 1, 1, 1)$
 \\ \hline
$E$   &
 $\left\{ \begin{array}{c}(0, -1, 1, -1, 1, 0)\\
(2, -1, -1, -1, -1, 2)\end{array}\right. $  &
$\left\{ \begin{array}{c}(-2, -1, -1, 1, 1, -2)\\
(0, 3, -3, -3, 3, 0)\end{array}\right. $
\\ \hline
$T_1$  & &
 $ \left\{ \begin{array}{c} (0, 0, -1, 0, -1, 0)\\
(0, 1, 0, 1, 0, 0)\\
(-1, 0, 0, 0, 0, 1)\end{array}\right. $
\\ \hline
$T_2$  &
 $ \left\{ \begin{array}{c}(0, 0, 1, 0, -1, 0)\\
(0, -1, 0, 1, 0, 0)\\
(-1, 0, 0, 0, 0, 1)\end{array}\right. $ &
\\ \hline
\end{tabular}
}\\
Table 12 Combination coefficients for the combination\\
 of $\{ \vv{a_i} \}$ from type 9.\\
\vspace*{0.4cm}

{\footnotesize
\begin{tabular}{|c|c|c|}\hline
~~~~ &  $PC=++$   &     $PC=--$  \\
     & $\{\vv{a}_i\},i=1,\cdots,6$  & $i=7,\cdots,12$   \\ \hline
$A_1$  &
$(1, 1, 1, 1, 1, 1)$ &
\\ \hline
$A_2$ &
$(1, -1, -1, -1, 1, 1)$ &
 \\ \hline
$E$   &
 $\begin{array}{c}
\left\{ \begin{array}{c}(-1, 1, 1, -2, 2, -1)\\
(-3, -3, 3, 0, 0, 3)\end{array}\right. \\
\left\{ \begin{array}{c}(1, -1, 2, -1, 1, -2)\\
(-3, -3, 0, 3, 3, 0)\end{array}\right. \end{array}$ &
\\ \hline
$T_1$  & &
 $ \left\{ \begin{array}{c} (0, 0, 0, -1, 0, -1)\\
(-1, 0, 0, 0, 1, 0)\\
(0, -1, 1, 0, 0, 0)\end{array}\right. $
\\ \hline
$T_2$  & &
 $ \left\{ \begin{array}{c}(0, 0, 0, -1, 0, 1)\\
(1, 0, 0, 0, 1, 0)\\
(0, 1, 1, 0, 0, 0)\end{array}\right. $
\\ \hline
\end{tabular}
}\\
Table 13 Combination coefficients for the combination\\
 of $\{ \vv{a_i} \}$ from type 10.\\
\vspace*{0.4cm}

{\footnotesize
\begin{tabular}{|c|c|c|}\hline
~~~~ &  $PC=++$   &     $PC=+-$  \\
     & $\{\vv{a}_i\},i=1,\cdots,3$  & $i=4,\cdots,6$   \\ \hline
$A_1$  &
$(1, 1, 1)$ &
\\ \hline
$E$   &
 $ \left\{ \begin{array}{c}(0, 1, -1)\\
(-2, 1, 1) \end{array}\right. $ &
\\ \hline
$T_1$  & &
 $ \left\{ \begin{array}{c} (0, 0, 1)\\(0, -1, 0)\\
(1, 0, 0)\end{array}\right. $
\\ \hline
\end{tabular}
}\\
Table 14 Combination coefficients for the combination \\
of $\{ \vv{a_i} \}$ from type 11.\\
\vspace*{0.4cm}

{\footnotesize
\begin{tabular}{|c|c|c|}\hline
~~~~ &  $PC=++$   &     $PC=+-$  \\
     & $\{\vv{a}_i\},i=1,\cdots,3$  & $i=4,\cdots,6$   \\ \hline
$A_1$  &
$(1, 1, 1)$ &
\\ \hline
$E$   &
 $ \left\{ \begin{array}{c}(0, 1, -1)\\
(-2, 1, 1)\end{array}\right. $ &
\\ \hline
$T_1$  & &
 $ \left\{ \begin{array}{c}(0, 0, 1)\\
(0, -1, 0)\\(1, 0, 0)\end{array}\right. $
\\ \hline
\end{tabular}
}\\
Table 15 Combination coefficients for the combination\\
 of $\{ \vv{a_i} \}$ from type 12.\\

\end{center}
\end{document}